\documentclass[aps,prl,twocolumn,showpacs,groupedaddress,amsmath,amssymb]{revtex4-2}

\usepackage{graphicx}
\usepackage{amsmath}
\usepackage{longtable}
\usepackage{xfrac}
\usepackage{color}
\usepackage{bm}
\usepackage[colorlinks,citecolor=blue]{hyperref}

\def\rbo{\mbox{\boldmath{$r$}}}
\def\rhobo{\mbox{\boldmath{$\rho$}}}

\def\xbo{\mbox{\boldmath{$x$}}}
\begin{document}

\title{ Possibility of forming a stable Bose-Einstein condensate of $2\,^{3}\!S_1$ positronium atoms}

\author{Y. Zhang$^{1,2}$}
 \author{M.-S. Wu$^{1}$}
  \author{ J.-Y. Zhang$^{1,3}$}\altaffiliation{Email address: jzhang@apm.ac.cn}
  \author{ Y. Qian$^{4}$}
   \author{X. Gao$^{3}$}
\author{K. Varga$^{5}$}
\affiliation{$^1$ State Key Laboratory of Magnetic Resonance and Atomic and Molecular Physics,
 Innovation Academy for Precision Measurement
Science and Technology, Chinese Academy of Sciences, Wuhan 430071, China}
\affiliation{$^2$ School of Physical Sciences, University of Chinese Academy of Sciences, Beijing 100049, China}
\affiliation{$^3$ Beijing Computational Science Research Center, Beijing 100193, China}
\affiliation{$^4$\,Department of Computer Science and Technology,
East China Normal University, Shanghai 200062, China}
\affiliation{$^5$ Department of Physics and Astronomy, Vanderbilt University, Nashville, Tennessee 37235, USA}

\date{\today}

\begin{abstract}
The confined variational method in conjunction with the orthogonalizing
pseudo-potential method and the stabilization method is used to
study the low energy elastic scattering between two spin-polarized metastable positronium  Ps(2\,$^{3}\!S_1$) atoms. Explicitly correlated Gaussian basis
functions are adopted to properly describe the complicated Coulomb
interaction among the four charged particles. The calculated $s$-wave
scattering length ($\approx8.5\,a_0$) is positive, indicating the
possibility of forming a stable Bose-Einstein condensate of fully spin-polarized
$\text{Ps}(2\,^{3}\!S_1)$ atoms. Our results will open a new way of
experimental realization of Ps condensate and development of $\gamma$-ray
and $\text{Ps}(2\,^{3}\!S_1)$ atom lasers.
\end{abstract}


\maketitle

The positronium (Ps) atom, a hydrogen-like bound system of an
electron and a positron, has two ground states,
the singlet 1\,$^{1}\!S_0$ state and the triplet 1\,$^{3}\!S_1$ state, known respectively as
 {\it para}-Ps ({\it p}-Ps) and {\it ortho}-Ps ({\it o}-Ps). {\it P}-Ps has a lifetime of 0.125~ns~\cite{dirac30} and decays into two gamma
photons, and {\it o}-Ps has a lifetime of
142~ns~\cite{ore1949a} and decays into three gamma photons. The longer lived {\it o}-Ps atoms were amongst the
first candidates {\cite{PhysRevB.49.454} for achieving Bose-Einstein condensation (BEC), a phase transition of a bose gas where a macroscopic number of bosons occupy a same quantum state below a
critical temperature.
BEC is one of the most interesting
phenomena in quantum systems of bosons that has important applications such as in
testing the weak equivalence principle and in studying gravitational effects on
quantum systems~\cite{edward2014a}. Historically, the first BEC was realized in an ensemble of Rb atoms in 1995~\cite{Anderson198}, which
opened a new era of ultracold physics. For {\it o}-Ps,
the smallness of its mass allows for much higher
BEC temperature of 20-30~K than ordinary atoms around 200~nK~\cite{anderson1995a,bradley1995a}.
However, a realization of BEC of {\it o}-Ps atoms has been hindered by its
very short lifetime.

The formation and observation of a Ps BEC has been of extraordinary interest though it is very challenging.
Since Platzman and Mills suggestion of a possible way of creating an {\it o}-Ps BEC in 1994~\cite{PhysRevB.49.454}, some significant progress has been made both theoretically and experimentally. Low energy scattering between two ground-state Ps atoms has been extensively studied for calculating scattering cross sections
and the $s$-wave scattering lengths~\cite{ivanov01b,shumway01a,oda2001a,ivanov02a,daily2015a}. These quantities are critical for determining possibility of forming a stable ground-state Ps BEC and for designing experimental configurations.
In order to probe Ps densities, some low-energy scattering properties of the ground- and $2s$-state of Ps have
been computed using hyperspherical coordinates~\cite{michael2019a}. For modeling a BEC process of {\it o}-Ps atoms confined in a porous
silica material, Morandi {\it et al.}~\cite{morandi2014a} showed that the condensation process is compatible with the {\it o}-Ps lifetime, which strongly depends on the
external electromagnetic field~\cite{cui2012a}.
There are also some theoretical works on $\gamma$-ray laser~\cite{avetissian14a,avetissian15a} and spinor dynamics~\cite{wang2014a,zheng17a} based on BEC of Ps atoms. From experimental side, significant progress has been made in the area of Ps-laser physics due to
the breakthrough development of the
Surko type buffer gas positron trap~\cite{surko1989a,murphy1992a,danielson2015a}.
Recently, implanting high density bursts of polarized positrons into a porous silica film in a high magnetic field, Cassidy
{\it et al.} produced
a highly spin-polarized $(96\%)$ {\it o}-Ps gas~\cite{PhysRevLett.104.173401}.
Moreover, the suppression of the Zeeman mixing
of the $2\,^1\!P$ and $2\,^3\!P$ states of Ps observed in high
magnetic fields~\cite{cassidy2011a} makes laser cooling of Ps feasible~\cite{LIANG1988419,iijima2001a,hirose2014a,shu2016a}.
Typically, Ps atoms produced in most porous materials can approach room temperature
at the currently highest achievable density $10^{16}~\text{cm}^{-3}$~\cite{PhysRevLett.104.173401}.
To form a Ps BEC, one needs not only increase the Ps density but also significantly reduce the temperature of Ps gas. However, the short lifetime of {\it o}-Ps seriously limits application of advanced cooling techniques, such as laser cooling, developed for ordinary atoms~\cite{Cassidy2018}. So far, {\it o}-Ps has been the only focus of all Ps-BEC related studies, although some experimental and theoretical researches have been conducted on the longer-lived metastable and Rydberg states of Ps~\cite{ziock1990a,estrada2000a,castelli2008a,cassidy2012a,PhysRevA.95.033408,PhysRevA.98.013402,amsler2019a,Cassidyb,antonello2019a}.

In this work, we will explore an alternative possibility of forming a BEC
using metastable $\text{Ps}^{*}(2\,^{3}\!S_1)$ atoms. In the following, the notation $\text{Ps}^{*}(2\,^{3}\!S_1)$ is abbreviated as
$\text{Ps}^{*}$. The $\text{Ps}^{*}$ has a
lifetime of 1136~ns that is eight times as long as the lifetime of {\it o}-Ps~\cite{ore1949a}.
By calculating the $s$-wave scattering length for the spin-aligned $\text{Ps}^{*}$-$\text{Ps}^{*}$ elastic scattering that governs the interaction between $\text{Ps}^{*}$ atoms at low temperatures, we will see whether it is a positive value, which is a key factor for forming a stable BEC.
$\text{Ps}^{*}$-atom scattering problem is one of the most difficult problems in atomic collision
theory because both projectile and target are composite objects with their internal
structures. One has to deal with multi-center integrals of interaction matrix elements.
A further complication to these calculations lies in the fact
that both colliding $\text{Ps}^{*}$ atoms are in the excited $2\,^{3}\!S_1$ state so that one should have basis functions to be able to accurately describe both short- and long-range (van der Waals) interactions, in particular for low energy scattering.

Based on the existing computational techniques~\cite{suzuki98a,zhang08b},
a novel method i.e. the confiend variation method (CVM) has been proposed recently~\cite{mitroy08d,zhang08c} for studying low-energy elastic scattering between a simple or composite projectile with an atom. The CVM combined with the orthogonalizing pseudo-potential (OPP)
method~\cite{krasnopol1974exclusion,PhysRev.153.177,PhysRevA.11.2018,MITROY1999103} will be applied to study the $s$-wave elastic scattering between two spin-aligned $\text{Ps}^{*}$ atoms.
The principal result of this work is that, for the first time, we have established a definitive value of the $s$-wave scattering length
that has positive sign, indicating that a stable BEC of spin-aligned $\text{Ps}^{*}$ atoms can be formed.

{\it Theory for the spin-aligned $\text{Ps}^{*}$-$\text{Ps}^{*}$ scattering}.\textemdash The nonrelativistic Hamiltonian for the four-body system of $(e^{+}e^{+}e^-e^-)$ can be written in the form (in atomic units)
\begin{eqnarray}
H &=& -\sum_{i=0}^{3} \frac{\nabla_{{\rbo}_i}^2}{2}+
\frac{1}{|{\rbo}_0-{\rbo}_1|}-
\frac{1}{|{\rbo}_0-{\rbo}_2|}-
\frac{1}{|{\rbo}_0-{\rbo}_3|}  \nonumber\\
&& -\frac{1}{|{\rbo}_1-{\rbo}_2|}-
\frac{1}{|{\rbo}_1-{\rbo}_3|}+
\frac{1}{|{\rbo}_2-{\rbo}_3|}  \,, \label {eq:cpt1}
\end{eqnarray}
where ${\rbo}_0$ and ${\rbo}_1$ are the two-positron position vectors, and ${\rbo}_2$ and ${\rbo}_3$ the two-electron position vectors.
For calculating the $s$-wave elastic scattering of $\text{Ps}^{*}$-$\text{Ps}^{*}$, we use the OPP
method~\cite{krasnopol1974exclusion,PhysRev.153.177,PhysRevA.11.2018,MITROY1999103} to prevent any electron-positron pair from forming the ground state $\text{Ps}(1\,^{3} \!{S}_1)$. The OPP operator is constructed by
summing over the $\text{Ps}(1\,^{3}\!{S}_1)$ projection operators
 \begin{eqnarray}
\lambda \hat{P}_{} &=& \lambda \sum^{1}_{i=0}  \sum^{3}_{j=2}\hat{P}_{ij} \\
&= & \lambda\sum^{1}_{i=0}  \sum^{3}_{j=2}|\phi_{1\,^{3}\!{S}_1}({\rbo}_i-{\rbo}_j)\rangle \langle \phi_{1\,^{3}\!{S}_1}({\rbo}_i-{\rbo}_j) | \, ,\nonumber
\end{eqnarray}
where $\lambda$ is a large positive number and $\phi_{1\,^{3}\!{S}_1}({\rbo}_i-{\rbo}_j)$ is the wave function of $\text{Ps}(1\,^{3}\!{S}_1)$.
Since a wave function with a nonzero overlap with the $\text{Ps}(1\,^{3}\text{S}_1)$ orbital tends to increase the energy, an eigenfunction of $ H +\lambda \hat{P}$ for a low energy level will have a very small overlap with $\phi_{1\,^{3}\text{S}_1}$.
The OPP method was first introduced by Krasnopolsky and Kukulin \cite{krasnopol1974exclusion} in 1974. Mitroy and Ryzhikh performed a comprehensive numerical investigation on the effects due to different $\lambda$ and different sizes of basis sets~\cite{MITROY1999103} and found that the energies calculated with OPP will converge to those of the $\hat{Q}\hat{H}\hat{Q}$ Hamiltonian in the projection operator method~\cite{Ryzhikh_1998},
a method that has been widely used in studying atomic and molecular resonant and excited states. Compared to the $\hat{Q}\hat{H}\hat{Q}$ method, the OPP method is easier to apply for scattering problems.

The $(e^{+}e^{+}e^-e^-)$ system possesses rich symmetries including the electron
interchange symmetry, the positron interchange symmetry, the inversion parity, and the charge parity. These symmetries can be described by a permutation group isomorphous to the molecular point group $D_{2h}$.
A detailed analysis of the symmetries is presented in Ref.~\cite{PhysRevLett.92.043401}.
For the fully spin-aligned $\text{Ps}^{*}$-$\text{Ps}^{*}$ scattering, the total spin operators $\hat{ \bf  S}^2$ and  $\hat{ \bf  S}_z$ have good
quantum numbers $S=S_z=2$, and this scattering state can be classified according to the irreducible representations of the $D_{2h}$ group as $B_1$ symmetry. In the calculation of the $s$-wave elastic scattering, even parities are used for both inversion and charge conjugation. After taking these symmetries into account, the total symmetry
projector applied to the spacial part of the basis function is $(1+P_{02} P_{13})(1-P_{01})(1- P_{23})$, where $P_{ij}$ is the permutation of the spatial coordinates of particles $i$ and $j$.

The Hamiltonian operator which is evaluated in the variational calculation usually commutes with all the permutation operators. Therefore we can perform a convenient implementation where all the permutational operators are applied to the ket
\begin{equation}
\langle\mathcal{P}\psi|\hat{H}|\mathcal{P}\psi\rangle=\langle\psi|\hat{H}|\mathcal{P}^{\dag}\mathcal{P}\psi\rangle,
\end{equation}
In the OPP method, The projection operators do not commute with all the permutation operators.
The total symmetry projector $\mathcal{P}=(1+P_{02} P_{13})(1-P_{01})(1- P_{23})$, adapting wave function to the correct $B_1$ symmetry actually, should be applied on both bra and ket. The matrix elements of Hamiltonian operator and OPP operator are written as
\begin{equation}
\hat{H}_{ij}+\lambda \hat{P}_{ij}=\langle\mathcal{P}\psi_i|\hat{H}+\lambda \hat{P}_{}|\mathcal{P}\psi_j\rangle,
\end{equation}
,which makes convergence rather slower.
Meanwhile the OPP operator $\lambda \hat{P}$ could result in linear dependence problems. Variational calculation became numerically unstable with respect to further enlargement of the ECG basis.

Of crucial importance to this work is the use of explicitly correlated Gaussians (ECGs)~\cite{boys60,singer60,cencek93a,suzuki98a} to describe the Coulomb interaction between the charged particles. An ECG basis can not only describe the correlations among the charged particles, but also allows us to
evaluate the Hamiltonian matrix elements analytically.
After separating out the center-of-mass motion from the $(e^{+}e^{+}e^-e^-)$ system, an ECG can be written in the form
\begin{equation}
\Phi_n =\exp{\left(-{1\over 2} \sum_{i\geq 1j \geq 1} A^n_{ij} {\xbo}_i \cdot {\xbo}_j \right)} \,,
\end{equation}
where $ {\xbo}_i=   {\rbo}_i-  {\rbo}_0 $.
The independent parameters $A^n_{ij}$ contained in symmetric matrices $A^n$ are optimized through the energy minimization using the confined variational method (CVM). The CVM is simple and powerful in the sense that it converts a problem of continuum states to a problem of bound states by adding a confining potential $\chi_{\rm cp}(\rho)$ to the Hamiltonian. This method provides a framework for optimizing wave functions in the interaction region using bound-state techniques. The advantages of using the CVM have been demonstrated by
solving some long-term intractable problems, including the $e^{+}\text{-H}_2$ scattering and $\text{Ps}\text{-H}_2$ scattering,
where the calculated annihilation parameters are, for the first time, in agreement with precise experimental values~\cite{zhang09b,zhang2018a}.
In this work, the confining potential is chosen to be
\begin{eqnarray}
\chi_{\rm cp}(\rho) &=& 0\, , \ \  \ \  \ \ \ \rho < R_0  \, ,\\
\chi_{\rm  cp}(\rho) &=& G (\rho - R_0)^2\, , \ \
\ \rho \geq R_0 \, , \label{ACP}
\end{eqnarray}
where $\rho$ is the distance between the two centers of mass
of two electron-positron pairs, and $G$ is a small positive number. We set
$ R_0 = 50\,a_0$ because the long-range interaction $V_{L}= -\text{C}_6/\rho^6$, where $\text{C}_6 = 27320~\text{a.u.}$~\cite{Zhangyia}, is required by the CVM to be weak at the boundary $R_0$. Due to the
exchange symmetries between the identical particles and their
indistinguishability, the Schr\"{o}dinger equation for the
confined $\text{Ps}^{*}$-$\text{Ps}^{*}$ system can be written in the form
\begin{eqnarray}
&& \Bigl[ H + \lambda \hat{P}_{}+\chi_{\rm cp}(\rho_{1}) + \chi_{\rm cp}\left(\rho_{2} \right)\Bigr]\Psi_i(\xbo) = E_i \Psi_i(\xbo)\,,
\label{schrod1}
\end{eqnarray}
where $\rho_{1} =|{\xbo}_1 +{\xbo}_{2}-{\xbo}_3|/2$, $\rho_{2} =|{\xbo}_1-{\xbo}_2 +{\xbo}_{3}|/2$, and $\xbo$ stands for $\{\xbo_{1},\xbo_{2},\xbo_{3}\}$ collectively.
Besides the ECG basis for the short-range interaction
region, as a supplement a set of exterior basis functions are designed to describe the long-range interaction between the two $\text{Ps}^{*}$, as listed below
\begin{eqnarray}
\Phi_{\text{ext}} = \text{exp}\bigl(-\frac{1}{2} \alpha_i \rho_1 \bigr)\phi_{\text{Ps}^{*}}({\xbo}_{3})\phi_{\text{Ps}^{*}}({\xbo}_{1}-{\xbo}_{2})\,,
\end{eqnarray}
where $\phi_{\text{Ps}^{*}}$ is the $\text{Ps}^{*}$
wave function written as a linear combination of 20 ECGs that give rise to an energy eigenvalue of $-0.062\,499\,999 \,999$~a.u. very close to  the exact value of -0.0625~a.u. for $2\,^{3}S_1$. A total of 10 even-tempered exponents $\alpha_i$ are generated using $\alpha_i=\alpha_1/T^{i-1}$ with $\alpha_1 =0.0001 $ and $T = 1.78$.

The phase shifts were extracted from wave functions using the stabilization method \cite{zhang08b}.
After omitting the confining potential in Eq~(\ref{schrod1}), the Schr\"{o}dinger equation was solved in the full interior and exterior regions to generate a set of positive energy pseudostates. The phase shift was derived by fitting the density distribution
$C({\rhobo})$ to the
asymptotical density distribution in the range $ \rho \in [ 40 a_0, 45 a_0]$, where $C({\rhobo})$ is defined as
\begin{eqnarray}
C({\rhobo})  =   &\int& d{\xbo}_1\,d{\xbo}_2\,d{\xbo}_3 \delta(({\xbo}_1+{\xbo}_2-{\xbo}_3)/2- \rhobo) \nonumber\\
&\times&  \left| \Psi({\xbo}_1,{\xbo}_2,{\xbo}_3)\right|^2\,.
\label{overlap}
\end{eqnarray}
To take into account the effect of the long range potential, the radial Schr\"{o}dinger equation of  the potential $V_L(\rho)$ scattering was numerically integrated inwards with the asymptotical wave function $ B \sin( k\rho + \delta_k)$.
Then after fitting the phase shift $\delta_k$ for small wave number $k$ to the effective-range expansion~\cite{drake06a}
\begin{eqnarray}
 k\cot(\delta_k)=-\frac{1}{A_0}+\frac{1}{2}r_0k^2 ,\label{eft}
\end{eqnarray}
one obtained the scattering length $A_0$.

{\it Results and discussion}.{\textemdash}
Table~\ref{scattB} presents the five lowest wave numbers, their corresponding phase shifts, and the determined $s$-wave scattering length at various stages of optimization for the inner basis functions. During the optimization, the OPP parameter $\lambda = 1000$ and the CVM parameter $G=4.56\times 10^{-3}$ were used in order to avoid linear dependence as we enlarged the size of basis set as large as possible.
%However, in order to obtain more accurate scattering parameters listed in Table~\ref{scattB},
%the largest $\lambda$ that we could choose without loss of precision was 5000, which was used for the
%stabilization calculations instead of $\lambda = 1000$.
As the size of basis becomes larger, the scattering length $A_0$ becomes larger.
\begin{table*}
\caption{Wave numbers $k_i$, corresponding phase shifts $\delta_{k_i}$, and determined $s$-wave scattering length $A_0$
for the fully spin-aligned $\text{Ps}^{*}$-$\text{Ps}^{*}$ elastic scattering, with the OPP parameter $\lambda = 1000$. In the first column, the
first entry is the dimension of the interior region basis and the second entry is for the exterior basis.
}\label{scattB}
\renewcommand\arraystretch{0.5}
\begin{tabular*}{\textwidth}{@{\extracolsep{\fill}}lccccc ccccc cc}
   \hline
   \hline
 $N$ & $k_1(a^{-1}_0)$  & $k_2(a^{-1}_0)$  & $k_3(a^{-1}_0)$ & $k_4(a^{-1}_0)$   & $k_5(a^{-1}_0)$ & $\delta_{k_1}(\text{rad})$  & $\delta_{k_2}(\text{rad})$ & $\delta_{k_3}(\text{rad})$  & $\delta_{k_4}(\text{rad})$  &
 $\delta_{k_5}(\text{rad})$  & ${A}_0(a_0)$  & ${r}_0(a_0)$\\
   \hline
3000+10 & 0.0402    & 0.0855   & 0.1327   & 0.1608   & 0.1842   & --0.3408  & --0.7942 & --1.385 & --1.781 & --2.042  & 8.0 & 6.1 \\
4000+10 & 0.0361    & 0.0768   & 0.1185   & 0.1312   & 0.1665   & --0.3043  & --0.7651 & --1.204 & --1.365 & --1.757  & 8.2 & 5.6 \\
5000+10 & 0.0330    & 0.0699   & 0.1085   & 0.1187   & 0.1520   & --0.2751  & --0.6823 & --1.094 & --1.235 & --1.469  & 8.3 & 5.3 \\
\hline \hline
\end{tabular*}
\end{table*}
Figure~\ref{shiftA} plots the phase shift $\delta_k$ for small wave numbers calculated with three
different values of $\lambda$. These lines represent the
effective-range fits to these data points obtained with the largest basis set composed of 5000 interior basis functions and 10 exterior basis functions. In addition, the $s$-wave scattering lengths $A_0$ obtained with different $\lambda$ are summarized in Table~\ref{scattA}. One can
see from Figure~\ref{shiftA} and Table~\ref{scattA} that, as $\lambda$ increases from 1000 to 5000, the overall accuracy of the phase shift and the scattering length improve.
From the trends of $A_0$ in Tables~\ref{scattB} and \ref{scattA}, we estimated the exact value of $A_0$ around $8.5 a_0$.
\begin{figure}
\includegraphics[scale=0.35]{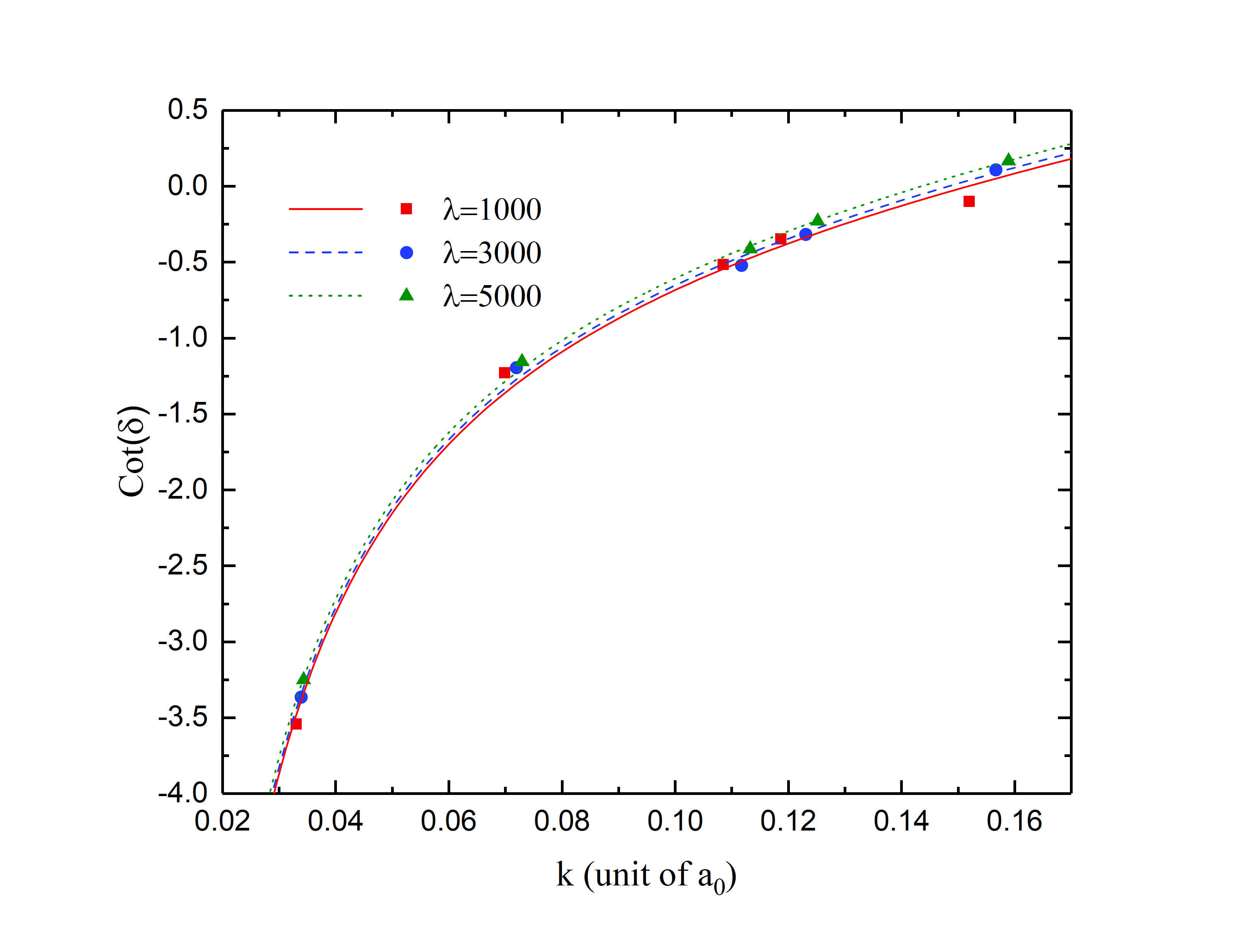}
\caption{$S$-wave phase shift $Cot(\delta_k)$ for the fully spin-aligned $\text{Ps}^{*}$-$\text{Ps}^{*}$ scattering as a function of wave number $k$ computed with three
different values of the OPP parameter $\lambda$.  The lines represent
effective-range fits to the phase shifts using Eq.~(\ref{eft}).}\label{shiftA}
\end{figure}
\begin{table}
\caption{$S$-wave scattering length for the fully spin-aligned $\text{Ps}^{*}$-$\text{Ps}^{*}$ elastic scattering obtained with different values of the OPP parameter $\lambda$. }\label{scattA}
\renewcommand\arraystretch{0.55}
\begin{tabular*}{8cm}{@{\extracolsep{\fill}}lccc}
   \hline
   \hline
  $\lambda$   & 1000  & 3000 & 5000 \\
 \hline
 ${A}_0(a_0)$ & 8.3   & 8.4  &  8.5  \\
 ${r}_0(a_0)$ & 5.3   & 5.4  &  5.7  \\
\hline \hline
\end{tabular*}
\end{table}

The positive value of $A_0$ means that in principle it is possible to produce a stable BEC of fully spin-aligned $\text{Ps}^*$ atoms.
For a low-density Bose gas, the interaction is dominated by the low-energy $s$-wave elastic scattering. Therefore, the $s$-wave scattering length plays an important role in the accurate description of static and dynamic properties of the low-density Bose gas.
For a gas of bosonic atoms trapped in a harmonic oscillator potential with an angular frequency $\omega_{\rm {HO}}$ and an oscillator length $a_{\text{HO}}=\sqrt{\hbar/(m\omega_{\text{HO}})}$, for example, the interaction is scaled by the ratio $A_0/a_{\rm HO}$.
For a $\text{Ps}^{*}$-BEC, this ratio becomes $A_0(\text{Ps}^{*})/a_{\rm HO} =2.25 (\text{eV})^{-1/2}\sqrt{\hbar\omega_{\text{HO}}}$, which is larger than $A_0(\text{{\it o}-Ps})/a_{\rm HO} = 0.81 \,(\text{eV})^{-1/2}\sqrt{\hbar\omega_{\text{HO}}}$ for $\text{{\it o}-Ps}$ \cite{ivanov01b}. Compared with $A_0(\text{H})/a_{\rm HO} =9.85 (\text{eV})^{-1/2}\sqrt{\hbar\omega_{\text{HO}}}$ for a polarized H~\cite{jamieson95a}, however, the interaction effect in $\text{Ps}^{*}$-BEC is obviously weaker and thus it is effectively closer to the ideal BEC than the H-BEC.
Using the value of $A_0(\text{Ps}^{*})/a_{\rm HO}$ , not only the Gross-Pitaevskii equation can be solved
to study the properties of the ground-state $\text{Ps}^{*}$-BEC but also the dynamics of phase transition can be investigated using the mean-field theory~\cite{RevModPhys.71.463}.

The necessary conditions for a realization of $\text{Ps}^{*}$-BEC are as follows.
Firstly, one should be able to produce a sufficiently high number of polarized $\text{Ps}^{*}$ atoms in a confined void. Secondly, the $\text{Ps}^{*}$ gas has to be cooled down to sufficiently low temperature. For cooling it can be achieved by thermalization through collisions of $\text{Ps}^{*}$ with the walls of the void, $\text{Ps}^{*} \text{-Ps}^{*}$ scattering, laser cooling, and other cooling methods. There has been a long interest in producing $\text{Ps}^{*}$~\cite{mills75a,chu1984a,fee1993a}
due to their potential applications~\cite{Cassidyb} in testing quantum electrodynamics (QED), in atom interferometry, and in gravitational interaction of antimatter. The available techniques of $\text{Ps}^{*}$ production include  radio frequency transition from laser-excited $\text{Ps}(2\,^{3} \text{P})$ in a weak magnetic field~\cite{mills75a}, two-photon Doppler-free $\text{Ps}(1\,^{3} \text{S})\text{-Ps}(2\,^{3} \text{S})$
laser excitation~\cite{chu1984a,fee1993a}, single-photon excitation of $\text{Ps}(1\,^{3} \text{S})$ to $\text{Ps}(2\,^{3} \text{P})$ in an electric field~\cite{PhysRevA.95.033408}, and radiative decay of $\text{Ps}(3\,^{3} \text{P})$ generated by single-photon excitation of $\text{Ps}(1\,^{3} \text{S})$~\cite{PhysRevA.98.013402, amsler2019a,antonello2019a}.
In particular,
the efficiency of $\text{Ps}^{*}$ production has recently been increased to $30\%$ by stimulating the $\text{Ps}(3\,^{3} \text{P})\text{-Ps}^{*}$ transition using a laser pulse, and further improvement in efficiency is still possible~\cite{antonello2019a}.
In addition, the highly polarized {\it o}-Ps gas and its corresponding production techniques~\cite{PhysRevLett.104.173401} will benifet to producing of highly polarized $\text{Ps}^{*}$ gas.
The produced $\text{Ps}^{*}$ can approach the room temperature 300~K after thermalization through collisions of $\text{Ps}^{*}$ with the walls of a void and through $\text{Ps}^{*} \text{-Ps}^{*}$ scattering. However, the density is five or more orders of magnitude lower than the required one at the corresponding temperature.
Therefore, it is necessary to further cool down the $\text{Ps}^{*}$ ensemble using laser cooling or other advanced cooling methods. So far $\text{ {\it o}-Ps}$ laser cooling has not been experimentally
realized due to the serious limitation of its short lifetime~\cite{LIANG1988419,iijima2001a,cassidy2011a,hirose2014a,shu2016a}.
A major advantage of using $\text{Ps}^{*}$ over {\it o}-Ps to realize a BEC is that $\text{Ps}^{*}$ has much longer lifetime than {\it o}-Ps.
However, laser cooling $\text{Ps}^{*}$ is very challenging and new cooling meathodologies  and technologies are required~\cite{LIANG1988419}.

The $\text{Ps}^{*}$-BEC can be applied  to study fundamental physics and create new technologies once it is formed. Ideally, it is possible to realize the transformation from $\text{Ps}^{*}$-BEC to {\it o}-Ps-BEC through the stimulated transition from $\text{Ps}^{*}$ to {\it o}-Ps. It is also possible to produce gamma-ray laser through the stimulated transition from $\text{Ps}^{*}$ to ${p}\text{-Ps}$ followed by the corresponding two-photon annihilation. Moreover, a coherent beam of $\text{Ps}^{*}$ atoms, the so-called $\text{Ps}^{*}$ atom laser, can be generated from a $\text{Ps}^{*}$ BEC.
Employing a $\text{Ps}^{*}$ atom laser as the $\text{Ps}^{*}$ source will significantly improve the accuracy of measurements on matter-antimatter gravitational interaction and on Ps precision spectroscopy~\cite{MILLS2002102,OBERTHALER2002129} which have been proposed to test QED and physics beyond the Standard Model~\cite{karshenboim2010a,shlomi2015a} such as the dark matter, and it will be of great benefit to producing cold antihydrogen atoms~\cite{bray2015a,mansouli2019a}. Furthermore, a coherent beam of $\text{Ps}^{*}$ atoms as a tool will enrich Ps chemistry to study various interactions with other atoms and molecules.

{\it Summary}.{\textemdash} In this work, the near-zero-energy $s$-wave elastic scattering between two fully-spin-aligned $\text{Ps}^{*}$  has been studied using a combined approach of OPP method, CVM, and stabilization method.
The calculated $s$-wave scattering length represents the
first determination of this quantity. The positive value of the scattering length ($\approx 8.5a_0$) is particularly significant since it demonstrates the feasibility of forming a stable BEC of fully-spin-aligned $\text{Ps}^{*}$ atoms and hence it becomes possible for developing $\gamma$-ray and  $\text{Ps}^{*}$ lasers based on $\text{Ps}^{*}$-BEC.

{\it  Acknowledgments}{\textemdash} J. Y. Z acknowledges S. Yi for valuable discussion and hospitality during his visit at the Institute of Theoretical Physics, Chinese Academy of Sciences. We would also like to thank Z.-C. Yan and W.-M. Liu for their helpful discussion. J. Y. Z. was supported by the Hundred Talents Program of the
Chinese Academy of Sciences. X. G. is supported by the National Natural Science Foundation of China (Grant Nos. 11774023 and U1530401), the National Key Research and Development
Program of China (Grant No. 2016YFA0302104).

\end{document}